\documentclass[aip,jap,reprint,amsmath,amssymb,superscriptaddress]{revtex4-1}
\usepackage{graphicx}
\usepackage{dcolumn}
\usepackage{bm}
\usepackage{color}

\begin{document}

\title{Controllable Spin-Charge Transport in Strained Graphene Nanoribbon Devices}
\author{Ginetom S. Diniz}
\email{ginetom@gmail.com}
\affiliation{Institute of Physics, University of Bras\'ilia, 70919-970, Bras\'ilia-DF, Brazil}
\author{Marcos R. Guassi}
\affiliation{Institute of Physics, University of Bras\'ilia, 70919-970, Bras\'ilia-DF, Brazil}
\author{Fanyao Qu}
\affiliation{Institute of Physics, University of Bras\'ilia, 70919-970, Bras\'ilia-DF, Brazil}
\affiliation{Department of Physics, The University of Texas at Austin, Austin, Texas 78712, USA}

\date{\today}

\begin{abstract}
We theoretically investigate the spin-charge transport in two-terminal device of graphene nanoribbons in the presence of an uniform uniaxial strain, spin-orbit coupling, exchange field and smooth staggered potential. We show that the direction of applied strain can efficiently tune strain-strength induced oscillation of band-gap of armchair graphene nanoribbon (AGNR). It is also found that electronic conductance in both AGNR and zigzag graphene nanoribbons (ZGNRs) oscillates with Rashba spin-orbit coupling akin to the Datta-Das field effect transistor. Two distinct strain response regimes of electronic conductance as function of spin-orbit couplings (SOC) magnitude are found. In the regime of small strain, conductance of ZGNR presents stronger strain dependence along the longitudinal direction of strain. Whereas for high values of strain shows larger effect for the transversal direction. Furthermore, the local density of states (LDOS) shows that depending on the smoothness of the staggered potential, the edge state of AGNR can either emerge or be suppressed. These emerging states can be determined experimentally by performing spatially scanning tunneling microscope or by scanning tunneling spectroscopy. Our findings open up new paradigms of manipulation and control of strained graphene based nanostructure for application on novel topological quantum devices.
\end{abstract}



\maketitle
\section{Introduction}
The atomically thick layer of carbon atoms with a honeycomb lattice structure - known as graphene - still keeps attracting considerable deal of attention due to its potential use in electronics. Graphene presents several exotic physical properties, such as the quantum spin Hall effect (QSHE) \cite{Kane,PhysRevLett.95.226801}, associated with a nontrivial topological time-reversal invariant state that has a bulk energy gap; and a pair of gapless spin filtered edge states at the sample boundaries. Recently, another striking topological phenomenon has attracted a notable interest, i.e. the quantum \emph{anomalous} Hall effect (QAHE) \cite{RevModPhys.82.1959}. This topological phenomenon appears if one of the spin channels in the QSH state is suppressed by the own sample magnetization, with the emergence of an electronic topological phase transition, characterized by quantized Hall conductance in an insulating state. This interesting phenomena predicted to be host in graphene \cite{PhysRevB.85.115439,PhysRevB.82.161414}, has been experimentally observed in topologically magnetic thin films \cite{Chang12042013}, with a great promise in spintronic applications \cite{RevModPhys.82.3045}.

A mechanism highly desirable in the development of spin-based devices is the effective control of the spin current flow \cite{Fabian}. To obtain polarized electronic current, one of the essential strategies is the creation of an effective potential barrier for a given spin specie, while the other can flow with no resistance. Also, the energy gap control, is an important issue related to the \emph{on-off} electrical tuning \cite{nl9039636}. One of the key elements to implement graphene based nanostructure in spintronic devices, relies upon the role of spin-orbit coupling (SOC). In graphene, there are two different SOC contributions: (i) the \emph{extrinsic} Rashba SOC, originated from the inversion symmetry breaking due to the substrate, which can also be manipulated by applied electric field \cite{Huertas,Sancho,Zarea} and (ii) the \emph{intrinsic} SOC (ISO) arising from the carbon atomic SOC - which is known to give rise to the existence of spin-polarized edge states in the QSH phase \cite{Haldane,Kane,PhysRevLett.95.226801,RevModPhys.82.3045}.

Mechanical deformations have a significant effect on the electronic, quantum transport and optical properties of a material and is used in the silicon electronics industry to boost device performance. For graphene and graphene nanoribbons (GNR) both experiments \cite{PhysRevB.79.205433,nn800459e,nl102123c,Levy30072010,PhysRevB.85.035422} and simulations \cite{PhysRevB.80.045401,PhysRevB.78.075435,PhysRevB.81.035411} have confirmed that the band structure can be dramatically altered by strain deformation. For instance, due to the breaking of sublattice symmetry, an uniaxial strain may induce a change of topology of the Fermi line, a merging of two inequivalent Dirac points and a tunable band gap at $K$ point \cite{PhysRevB.80.045401,PhysRevB.81.035411}. Thus, it may be used to trigger a quantum phase transition from a semi-metal to a semiconductor.

A combination of basic elements of uniaxial strain, SOC, exchange field, and staggered potential leads to a very exotic physics \cite{Diniz}. For instance, intrinsic SOC is favorable for opening a bulk energy gap around Dirac points \cite{Haldane,Kane,PhysRevLett.95.226801}, while Rashba SOC depresses the gap \cite{PhysRevB.81.165410,Zarea}. Then intrinsic and Rashba SOCs make an opposite contribution to the topological phase of QSHE. Besides, Rashba SOC breaks down inversion symmetry and exchange term breaks down time-reversal symmetry (TRS). Thus changing RSO and exchange interaction may lead to a QSH to QAH phase transition \cite{PhysRevB.85.115439,PhysRevB.82.161414}, which can be further manipulated by the application of uniaxial uniform strains \cite{Diniz}. Although, the quantum conductance of GNRs under uniaxial strain has been previously reported \cite{PhysRevB.81.024107,Nanoresearch,Wang}, the spin related terms and staggered potential were not taken into account.

In this paper, we aim to analyse the electronic transport control and the LDOS in GNR with different terminations in the QAHE phase by means of uniform strain deformations and smoothness of staggered potential. The electronic transport can be performed by using a two-terminal device akin to a field electron transistor (FET). QAHE phase can be determined experimentally, by on-site spin-resolved density of states, that can be accessed by spatially scanning tunneling microscope (STM) or by scanning tunneling spectroscopy (STS) \cite{Morgenstern,Stolyarova29052007,Eva}.

%
%
%
\section{Theoretical Model}
We consider GNR with homogeneous SOCs, exchange field interaction and staggered potential. The uniaxial strain is included through the introduction of strain-dependent hopping parameters. The Hamiltonian for this system in the real space reads
\begin{eqnarray}
\label{H1}
H = H_{0} + H_{R} + H_{ISO} + H_{M} + H_{U},
\end{eqnarray}
where $H_{0}=\sum_{\langle i,j\rangle,\sigma}t_{ij}c_{i\sigma}^{\dagger}c_{j\sigma}$ is the nearest-neighbor $\pi$-band tight-binding Hamiltonian. The fermionic operators $c_{i\sigma}^{\dagger}$/$c_{i\sigma}$ creates/anihilates an electron at site \emph{i} with spin $\sigma$ ($=\uparrow, \downarrow$) and hopping amplitude $t_{ij}=t_{i}=t_{0}e^{-3.37\left(\delta_{i}-1\right)}$. The unstrained $t_{0}\approx 2.7$eV \cite{PhysRevB.80.045401} and the deformed lattice distances $\vec{\delta}_{i}$ are related to the relaxed ones $\vec{\delta}_{i}^{0}$ by $\vec{\delta}_{i}=\left(1+\epsilon\right)\vec{\delta}_{i}^{0}$, where the uniaxial strain tensor is given in ref. \cite{PhysRevB.80.045401}. For simplicity, we set the unstrained C-C distance to be unity. $H_{R}=\sum_{\langle i,j\rangle,\sigma} i\lambda_{R}\sum_{\langle i,j\rangle}\hat{z}\cdot (\vec{s}\times \vec{\delta}_{ij})c_{i}^{\dagger}c_{j}$ is the Rashba spin-orbit with interaction with parameter $\lambda_{R}$ proportional to the electric field applied perpendicular to the $x$-$y$ plane of the graphene \cite{PhysRevLett.95.226801,Zarea,Huertas}, $\vec{s}$ are the Pauli spin matrices. $H_{ISO}=(2i/\sqrt{3})\lambda_{so}\sum_{\langle\langle i,j\rangle\rangle}c_{i}^{\dagger}\vec{s}\cdot \left(\vec{d}_{kj}\times \vec{d}_{ik}\right)c_{j}$ is the intrinsic SOC, and the vectors $\vec{d}_{ij}$ points from site $j$ to $i$, which for the intrinsic SOC with coupling parameter $\lambda_{so}$, which connects the next nearest-neighbors through $k$ \cite{Haldane,Kane,PhysRevLett.95.226801}. $H_{M} = M\sum_{i;\sigma,\sigma^{\prime}}c_{i\sigma}^{\dagger}s_{\sigma\sigma^{\prime}}^{z}c_{i\sigma^{\prime}}$ corresponds to the uniform exchange field with strength $M$, responsible for breaking TRS of the system \cite{PhysRevB.83.155447} and $H_{U} = \sum_{i;\sigma}V_{i}c_{i\sigma}^{\dagger}c_{i\sigma}$ refers to the staggered sublattice confining potential with $V_{i}=\pm V_{0}[e^{-(y_{i}/\xi)} + e^{-(y_{N_{A(Z)}-}y_{i}/\xi)}]$, $\pm$ being for $A/B$ sublattice with value $\pm V_{0}$ at the edges \cite{PhysRevLett.108.196806}. The $V_{i}$ is strongly dependent across the transversal direction, and decays exponentially from the edges with a characteristic width $\xi$.

\begin{figure}[!h]
\centering
\includegraphics[scale=0.42]{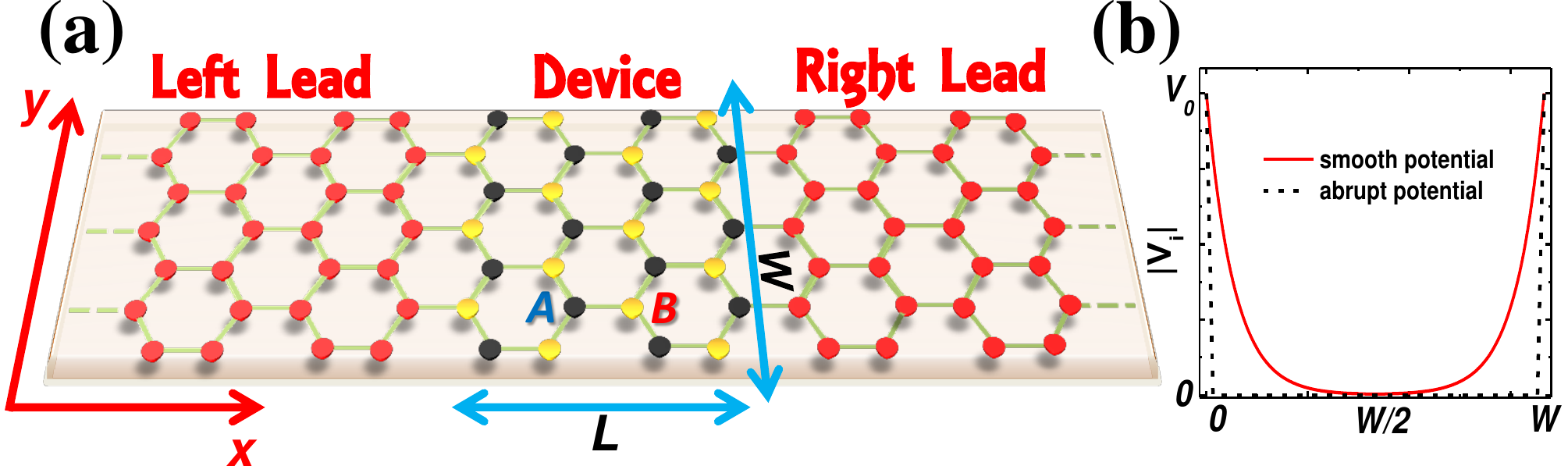}
\caption{(a) Schematic picture of a two terminal electronic AGNR device connected to semi-infinite left/right leads, with $L=(3/2N_{Z}-1)$ and $W=\sqrt(3)/2(N_{A}-1)$. (b) Potential profiles of $\vert V_i\vert$ along the transversal direction, for an abrupt confining potential (dashed line) and a smooth confining potential (solid line).}
\label{grafico1}
\end{figure}

To calculate the spin-resolved conductance and the LDOS, we have implemented the standard surface Green's function approach \cite{Nardelli,Sancho1,Sancho2}. The GNR device is divided into three regions: left lead, central conductor and right lead, as schematically shown in Fig. \ref{grafico1} (a) for an AGNR.  Notice that $N_{Z}$ denotes the number of zigzag chains and $N_{A}$ to the number of dimer lines. The uniaxial strain is applied to either the longitudinal ($\theta$=0) or the transversal ($\theta$=$\pi/2$), or at any arbitrary angle with respect to the $x$-axis. The central conductor is the only region under the influence of SOC effects, exchange field and staggered potential; it is also connected to semi-infinite leads by nearest-neighbor hopping. To avoid surface mismatch \cite{PhysRevB.88.195416}, we have considered that the leads are also strained. Therefore, a perfect atomic matching at the interface leads/central conductor is achieved. The Green's function of the device (omitting the spin indices) is then calculated by
\begin{equation}
\mathcal G_{C}^{a/r}(E)=\left(E \pm i\eta -H_{C}-\Sigma_{L}-\Sigma_{R}\right)^{-1},
\end{equation}
where $a/r$ denotes the advanced/retarded Green's function, $E$ is the energy ($\eta\rightarrow 0$) of the injected electron (the Fermi energy at a given doping). $H_{C}$ stands for the Hamiltonian in the central region and $\Sigma_{L/R}$ are the self-energies that describe the influence of the left/right leads, $\Sigma_{l}=H_{lC}^{\dagger}g_{l}H_{lC}$, where $g_{l}$ is the Green's function for the $l=L,R$ semi-infinite lead, obtained through an iterative procedure of the tight-binding Hamiltonian \cite{Nardelli}, and $H_{lC}$ couples each lead to the central region. The spin resolved conductance through the system is given by,
\begin{equation}
G_{\sigma\sigma^{\prime}}=G_{0}Tr\left[\Gamma_{\sigma}^{L}\mathcal{G}_{C,\sigma\sigma^{\prime}}^{r}\Gamma_{\sigma^{\prime}}^{R}\mathcal{G}_{C,\sigma^{\prime}\sigma}^{a}\right],
\end{equation}
where the trace runs through the lattice sites at the central conductor, $G_{0}=e^2/h$ is the quantum of conductance per spin, and $\Gamma_{\sigma}^{l}$ are the couplings for the leads, related to the spin-diagonal self-energies by $\Gamma^{l}=i\left[\Sigma_{l}^{r}-\Sigma_{l}^{a}\right]$ \cite{Nardelli}.

\section{Results}
In what follows, we focus on a ZGNR ($N_{Z}=26$) with a specific width of 5.4nm and length of 10.93nm (while in absence of strain). Similarly, we have chosen a metallic AGNR, i.e $N_{A}=3N-1$, with $N$ being a positive integer, with approximately same width and length. It is important to mention that wider the ribbon, more conducting channels can be available for higher doping, therefore the problem becomes more complex for both ZGNR and AGNR. Although, close to the Fermi level there will be the same amount of conducting channels for the pristine metallic nanoribbons. We have also checked the length dependence, and it is indeed \emph{relevant} to the device prototype, as the scattering region, i.e. the region of the device has the effects of SOC, exchange field and staggered potential enhanced, as the injected electron can \emph{feel} such fields in a longer length scale. Although we choose a specific length, the results presented here show a general behavior of such devices.

%
\begin{figure}[!h]
\centering
\includegraphics[scale=0.52]{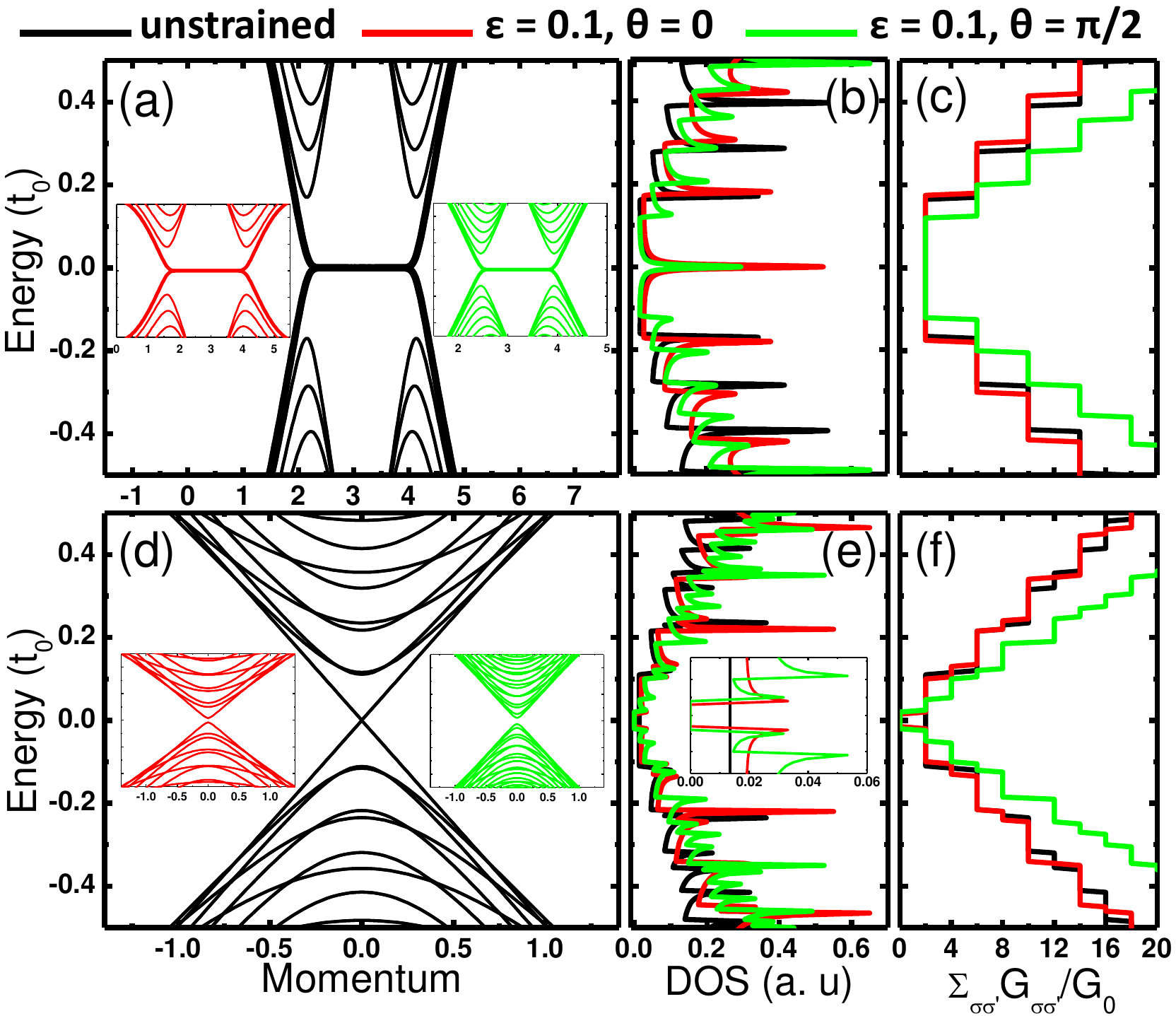}
\caption{Electronic structure of unstrained ZGNR with $N_{Z}=26$ (a) and AGNR with $N_{A}=47$ (d) with only $\pi$-band tight-binding Hamiltonian. The insets in panel (a) and (d) depict the correspondent strained GNR with $\varepsilon = 0.1$, and $\theta=0$ (red) and $\theta=\pi/2$ (green) directions. (b)-(c) and (e)-(f) Correspond to the density of states and the total conductance associated with panel (a) and (d), respectively. The inset in (e) shows clearly the vanishing of the density of states close to the Fermi level for the strained AGNR device.}
\label{grafico2}
\end{figure}
To illustrate the band structure of pristine GNR under uniaxial strains, we assume that except for the first term all other terms in Eq. \ref{H1} are set to zero. Fig. \ref{grafico2} (a) and (d), show the electronic structure of unstrained and strained (insets) ZGNR and AGNR with nearest-neighbors hopping, while (b) and (e) plot their correspondent density of states. In contrast, (c) and (f) illustrate the conductance profiles. One can notice the edge states of ZGNR (thick solid line) at the Fermi level for 26-ZGNR, is robust against uniaxial strains, as demonstrated in panel (b) and (c). The ribbon is metallic due to the strain-robust zero-energy flat band, and the conductance turns out to be an integer multiple of the quantum conductance $G_{0}$, as the channels are completely transparent. For high energy doping, there is only a slightly change in the plateaus, and as the ZGNR has two distinct valleys, the conductance increases always by 4$G_{0}$.

In strong contrast, the band structures of the AGNR families (semiconducting or metallic) are highly sensitive to the tensile uniaxial strain. As strain increases, the electrical conductivity may change from an insulator to a conductor or vice-versa. But, the prominent effect is observed near the Fermi level, as strain can induce a band gap in an otherwise metallic AGNRs, or change an existent band gap in semiconducting ribbons in a non-monotonic way \cite{Nanoresearch}. This band gap can be explained by the strain-induced shifting of the Dirac point, similar to previous works on deformed carbon nanotubes \cite{Yang}. For instance, strains drastically change the electronic behavior of metallic 47-AGRN, shown in (d), responsible for an electronic topological transition (ETT) from metallic to semiconductor. It manifest itself for the appearance of a vanishing density of states close to the charge neutrality point (highlighted in the inset of panel (e)). In addition, the transport gap ($\Delta_{g}$) shows a strain direction dependence, which has been intensively explored by other authors \cite{Nanoresearch,PhysRevB.81.205437,PhysRevB.81.125409,PhysRevB.81.193404}, which demonstrated that is robust even across strained junctions or intrinsic defects \cite{PhysRevB.88.195416,PhysRevB.88.125420}.

\begin{figure}[!h]
\centering
\includegraphics[scale=0.50]{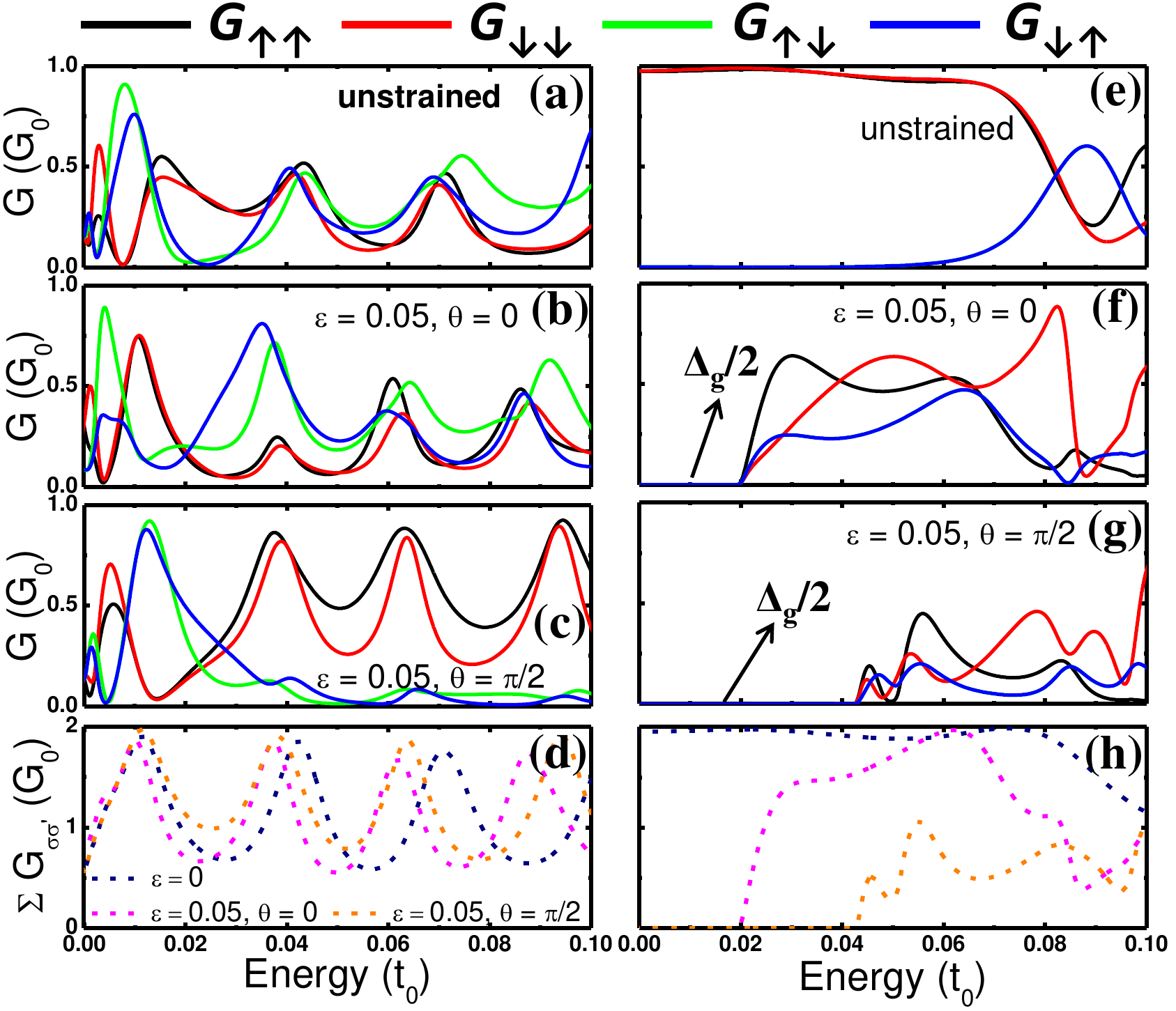}
\caption{Effects of strain on spin resolved conductance $G_{\sigma\sigma^{\prime}}$ of ZGNR with $N_{Z}$=26 (a)-(c) and AGNR with $N_{A}$=47 (e)-(g), respectively. Panels (d) and (h) show the total conductance. The parameters used in all panels are $\lambda_{R}$=0.1$t_{0}$, $\lambda_{so}$=0 and $M$=$0.2t_{0}$.}
\label{grafico3}
\end{figure}

To study the conductance characteristics in presence of both Rashba SOC and exchange field, we set the parameters $\lambda_{R}$=0.1$t_{0}$, $M$=0.2$t_{0}$, and $\lambda_{so}$=0$t_{0}$. Notice that with these parameters, the system is in the QAHE phase \cite{PhysRevB.85.115439}. Nevertheless, if the ISO parameter is different from zero, there is an upper-limited value of $\lambda_{so}$ \cite{PhysRevB.85.115439,Diniz}, beyond which a new phase characterized by a vanishing Chern Number $\mathcal{C}$=0 can take place; this phase is the so-called TRS-broken QSH phase \cite{PhysRevLett.107.066602,PhysRevB.85.115439,Diniz}. The spin-resolved conductance $G_{\sigma\sigma^{\prime}}$ is shown in Fig. \ref{grafico3}: for (a) unstrained, (b) strained along $\theta=0$, (c) strained along $\theta=\pi/2$ and (d) the total conductance $\sum_{\sigma\sigma^{\prime}} G_{\sigma\sigma^{\prime}}$ of a ZGNR. Notice that there is a suppression for both the spin conserving and the spin-flip conductance components for either unstrained or strained ZGNR in the energy range considered. However, these backscattering (transmission dips) at certain precise energies at the first plateau are different depending on the strain configuration, and a close inspection shows that conducting channels for non-spin flip and spin-flip conductances oscillate. Depending on the Fermi energy and set parameters, certain conductance components can even be completely suppressed. This suppression is attributed to the appearance of quasi-localized states in the device, which may produce sharp scattering resonances, also known as resonant backscattering which is a general behavior of quasi-1D quantum systems \cite{PhysRevLett.86.4275}. For higher energies, however, the large number of conducting channels leads to a non-vanishing transmission, as the channels get mixed along the device, and results in the appearance of an interchannel backscattering dominated by interference effects. Therefore, in the QSH phase protected by the TRS, nonmagnetic impurities do not cause backscattering on each boundary, and the spin transport in the edge states is dissipationless at zero temperature. In the QAH phase, however, there is a weak scattering between forward and backward movers, leading to a low-dissipation spin transport. At low energy, this interesting strain-controllable behavior of conducting channel suppression might be efficiently used to filter electrical current of desired spins, in spin filtering devices. In Fig. \ref{grafico3} (d), we show the total conductance, which is nearly robust against strains, specially close to the charge neutrality point, where the deviations due to strain are quite small. In contrast, the conductance of AGNR shows a drastic modification, with the development of a transport gap, which is insensitive to the electron spin that is injected-collected in the device. However, this induced transport gap is dependent upon the direction of the applied strain, with a larger conduction suppression along $\theta=0$ (red dashed line) with $\Delta_{g}$= 0.04 $t_{0}$, and $\Delta_{g}$= 0.086 $t_{0}$ while along $\theta=\pi/2$, that can be observed in panel (h). Also, the total conductance exhibits different plateaus: around $2G_{0}$ and approximately $G_{0}$ in AGNR without and with strain, respectively, which is one less quantum of conductance available for the electron to be transmitted along the device.

\begin{figure}[!h]
\centering
\includegraphics[scale=0.55]{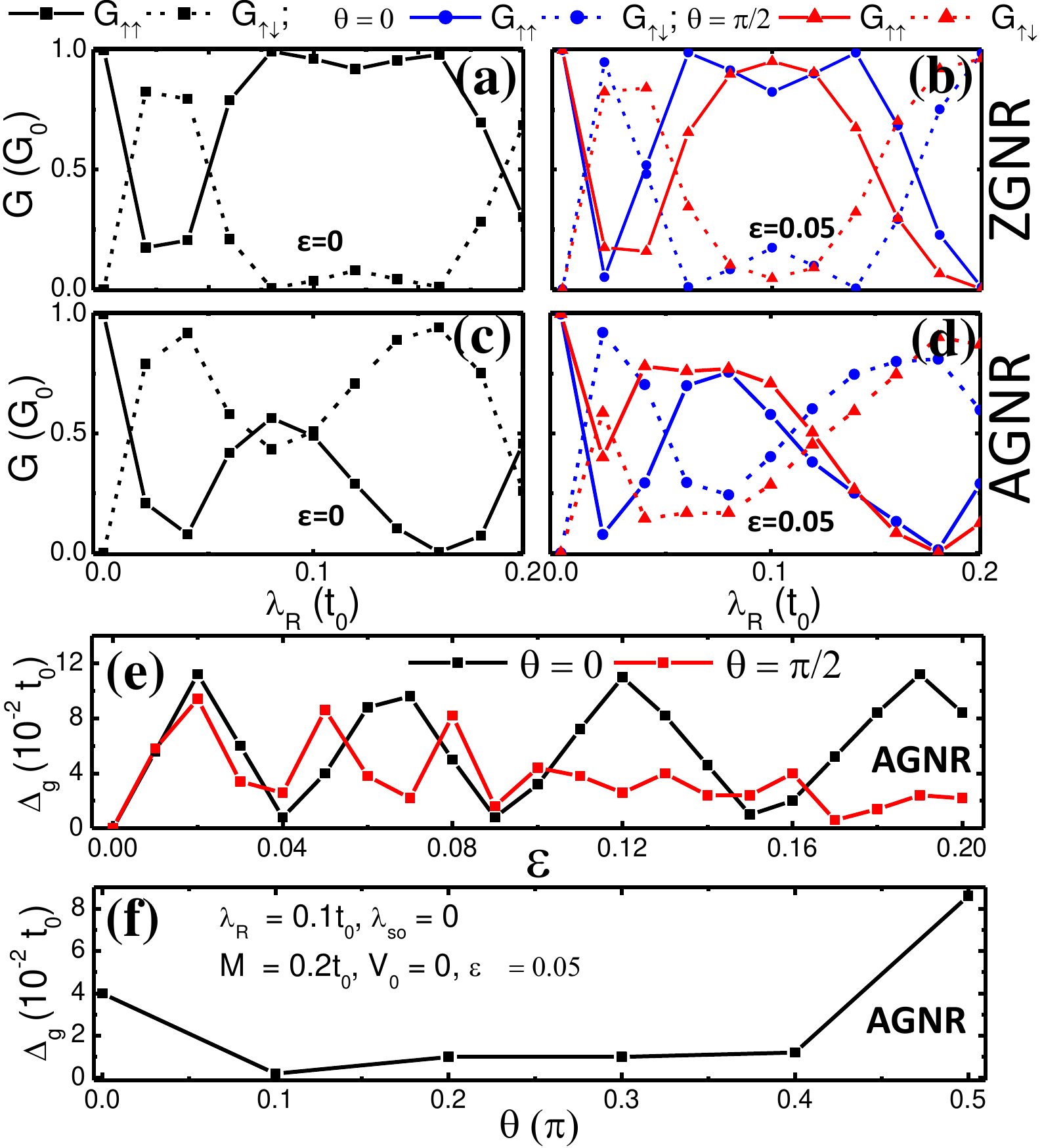}
\caption{(a) Conductance $G_{\uparrow\uparrow}$ and $G_{\uparrow\downarrow}$ for 26-ZGNR (a)-(b) and 47-AGNR (c)-(d) as function of $\lambda_{R}$ subjected to different configurations of strain. Panels (e) shows the band gap $\Delta_{g}$ of an AGNR as function of strain parameter $\varepsilon$ for $\theta=0$ and $\theta=\pi/2$, respectively. (f) $\Delta_{g}$ of an AGNR as function of the direction of strain for fixed $\varepsilon$=0.05.}
\label{grafico4}
\end{figure}

Another remarkable phenomenon is the oscillatory dependence of the spin components of $G_{\sigma\sigma^{\prime}}$ on the value of $\lambda_R$, which is shown in Fig. \ref{grafico4} (a)-(d), where the curves correspond to different topological GNRs and strain setups for $E=0.05t_{0}$.
The same parameters are used as the Fig. \ref{grafico3}, except for $M$. To reveal the effects of Rashba SOC, we set $M$=0 in the calculation. Then, the system is time-reversal invariant and the conductance components $G_{\uparrow\uparrow}=G_{\downarrow\downarrow}$ and $G_{\uparrow\downarrow}=G_{\downarrow\uparrow}$. This oscillatory behavior is reminiscent of the spin field effect transistor (FET) and has a similar source \cite{Datta-Das}, as the spin {\em precesses} as it propagates in the presence of the Rashba field, acquiring a net phase that is proportional to $\lambda_{R}L$, where $L$ is the length of the device. Further inspecting the strain-induced band gap in 47-AGNR in presence of SOC and exchange field interactions, one notices that in Fig. \ref{grafico4} (e) a similar band gap oscillation characteristic as reported in a earlier work \cite{Nanoresearch}. In the regime of small strain, the band gap shows approximately linear response, with increasing values of strain, however, it starts to oscillate. Further investigation shows that the amplitude and period of the gap-oscillation are tuned by direction of the strain, as shown in Fig. \ref{grafico4} (e). A specific dependence of transport-gap on the angle of the strain is clearly depicted in Fig. \ref{grafico4} (f). Notice that the transport gap is indeed strongly tuned by strain-direction. It equals approximately zero at 0.1$\pi$, while it reaches 0.086 $t_{0}$ at 0.5$\pi$.

\begin{figure}[!h]
\centering
\includegraphics[scale=0.8]{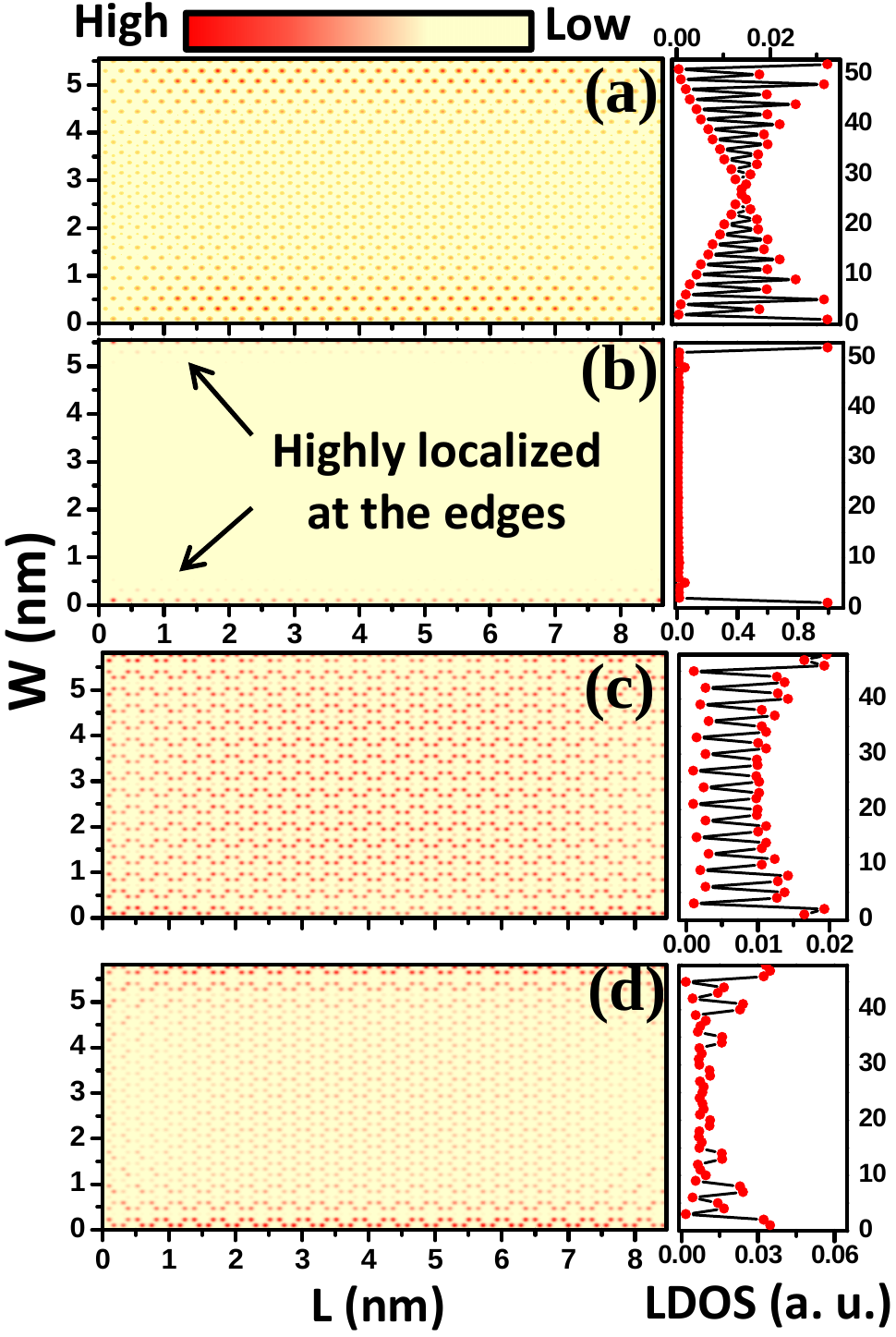}
\caption{Calculated STM maps for ZGNR with $N_{Z}$=26 for the $down$ (a) and $up$ (b) spins, respectively. STM maps for AGNR with $N_{A}$=47 for the $down$ (c) and up (d) spins. The panels on the right show the LDOS along the transversal direction close to the leads contact. The parameters used are $\lambda_{R}$=0.1$t_{0}$, $\lambda_{so}$=0.05$t_{0}$ and $M$=$0.2t_{0}$.}
\label{grafico5}
\end{figure}

%
Fig. \ref{grafico5} (a)-(d) show the theoretical STM maps for an incoming electron with Fermi energy of $E$=0.05$t$ for $\lambda_{R}$=0.1$t_{0}$, $\lambda_{so}$=0.05$t_{0}$ and $M$=$0.2t_{0}$. The panels on the right show the LDOS across the transversal direction.  These STM maps can be experimentally accessible by performing a STM or through a STS measurements. The calculation has been performed by using the LDOS, $\rho_{ii}=-\pi^{-1} Im(\mathcal G_{C}^{r}(E)_{ii})$, and with the aid of a $\pi$-atomic orbital to smooth out the STM maps. It is important to mention that the system is in the QAHE, with a non-zero Chern number \cite{PhysRevB.85.115439}. Fig. \ref{grafico5} (a)-(b), show the highly localized edge states present in ZGNR in the QAHE phase. Breaking of the TRS due to the exchange field, leads to a suppression for the spin down states in (a). Nevertheless, for the spin up there is an emerging edge localization as shown in (b). In contrast, no edge localization is found in AGNR as shown in Fig. \ref{grafico5} (c)-(d). Therefore, the conductance in AGNR is achieved by the bulk conducting channels.

\begin{figure}[!h]
\centering
\includegraphics[scale=0.8]{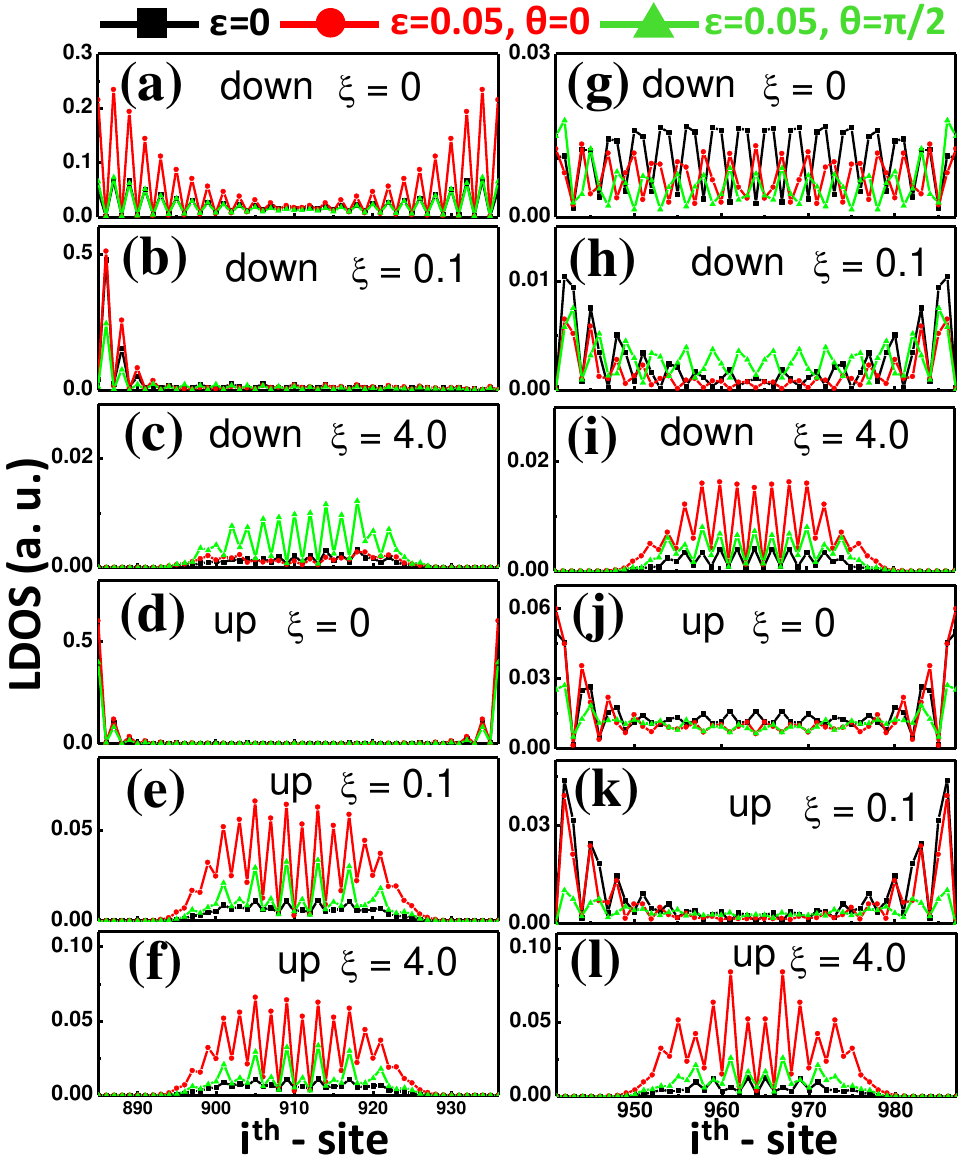}
\caption{LDOS along the width for 26-ZGNR (a-f panels) and 47-AGNR (g-l panels) at different smoothness parameters and spin states, with staggered potential $V_{0}$= 6$t_{0}$. Here, $\lambda_{R}$=0.1$t_{0}$, $\lambda_{so}$=0.05$t_{0}$ and $M$=$0.2t_{0}$ have been used. The curves are for different configurations of applied strain in the system.}
\label{grafico6}
\end{figure}

To further inspect the edge state behavior and its robustness against a smooth staggered sublattice potential, we have calculated the LDOS for different parameter configurations for both 26-ZGNR and 47-AGNR for fixed $E$=0.05$t_{0}$, with parameters $\lambda_{R}$=0.1$t_{0}$, $\lambda_{so}$=0.05$t_{0}$ and $M$=$0.2t_{0}$. In Fig. \ref{grafico6} (a)-(f), we show the LDOS across at exactly the middle of the device, i.e. at a length $L/2$, of the 26-ZGNR while (g)-(l) corresponds to 47-AGNR, for different set of parameters. For 26-ZGNR, we can notice that the LDOS is robust against strain with a slight modification on the amplitudes of the LDOS. However, when a staggered sublattice potential is considered by setting $V_{0}$=6$t_{0}$, smoothness parameter $\xi$ dependent behavior emerges. For instance, by setting $\xi$=0.1 implies in the suppression of the LDOS in one of the edges for a given spin. Also, there is a clear asymmetry, that can be explained by the A/B sublattice asymmetry - in one side, the edge terminates in A-site, while in the other in B-site. This characteristic makes the staggered potential being an effective barrier - indeed for both spin species when $\xi$=4.0 is set - as the staggered potential can be considered as a non-magnetic impurity. Also, the LDOS amplitudes has some important changes while applying strain with $\varepsilon=0.05$ along longitudinal ($\theta=0$) and transversal direction ($\theta=\pi/2$), which can be understood by compression/elongation of the ribbon width that is directly associated to the length smoothness of the staggered potential.

For the case of an AGNR, away from strain induced transport gap $\Delta_{g}$ (in fact it does not change the strain induced transport gap), we can also observe changes in the LDOS amplitude for different strain configurations. However, some important aspects can be observed by adding a staggered potential: (i) - for $\xi=0.1$, that would correspond to a less smooth potential (the potential effectively zero in the central region of the ribbon), there is an emerging localization in the AGNR akin to the edge states usually observed in ZGNR \cite{PhysRevLett.108.196806}, and perfectly symmetric on both edges as it does have same sublattice termination. (ii) for $\xi=4.0$, that would correspond to smother staggered potential, the edge localized LDOS is fully suppressed for both spin species, and the contributing conducting channels are now at the central region. So the transition from an abrupt to a smoother staggered potential might be traced down to a topological phase transition with the quenching of an emerging edge state, therefore the Chern number is expected to vanish in this condition \cite{PhysRevB.85.115439,Diniz}.

\section{Conclusions}
In summary, we have investigated the spin-resolved electronic transport and LDOS of GNR devices under the influence of SOC, exchange field, smooth staggered potential and uniform uniaxial strains. Our results demonstrate that it is possible to achieve a total electron transmission suppression of specific spin specie, which can be further tailored by uniaxial tensile strain on specific directions. Furthermore, by including a graded staggered potential, the following interesting behaviors have been observed in the LDOS maps: (i) selective edge conducting channel suppression for ZGNR for a sharper staggered potential and (ii) emerging of an edge state for AGNR for sharper staggered potential, which is associated to a topological phase transition \cite{PhysRevLett.108.196806}. These results suggest a possible implementation of a field-effect topological quantum transistor based on strained GNR, thus paving the way for the development of novel topological quantum devices.

\acknowledgements
We thank discussions with M. Ezawa, Shu-Shen Li and Z. Qiao. We acknowledge financial support received from CAPES, FAP-DF and CNPq.


\end{document}